\documentclass[format=acmsmall,review=False,screen=true]{acmart}
\usepackage{amssymb}
\usepackage{hyperref}
\usepackage{bm}
\usepackage{multirow}
\usepackage{ulem}
\usepackage{booktabs} 
\usepackage{graphicx}
\usepackage{tabularx}
\usepackage{diagbox}
\usepackage{graphicx}
\usepackage{subfigure}
\usepackage{float}
\usepackage{color,xcolor}
\usepackage[ruled]{algorithm2e} 
\usepackage{threeparttable}
\usepackage{soul}

\SetAlFnt{\small}
\SetAlCapFnt{\small}
\SetAlCapNameFnt{\small}
\SetAlCapHSkip{0pt}
\IncMargin{-\parindent}
\sethlcolor{yellow} 
\definecolor{gold_yellow}{RGB}{255,255,15}
\definecolor{orange_yellow}{RGB}{255,162,47}
\AtBeginDocument{%
  \providecommand\BibTeX{{%
    \normalfont B\kern-0.5em{\scshape i\kern-0.25em b}\kern-0.8em\TeX}}}

\begin{document}

\setcopyright{acmcopyright}
\acmJournal{TOIS}
\acmYear{2019} \acmVolume{1} \acmNumber{1} \acmArticle{1} \acmMonth{1} \acmPrice{15.00}\acmDOI{10.1145/3361719}

\title[Hiepar-MLC for paper-reviewer recommendation]{A multi-label classification method using a hierarchical and transparent representation for paper-reviewer recommendation}

\thanks{*Corresponding Author: Shu Zhao(zhaoshuzs2002@hotmail.com)}
\author{Dong Zhang}
\affiliation{%
  \institution{Anhui University}
  \department{Key Laboratory of Intelligent Computing and Signal Processing of Ministry of Education}
  \department{School of Computer Science and Technology}
  \city{Hefei}
  \postcode{230601}
  \country{China}
}
\email{darkzd_zhang@163.com}
\author{Shu Zhao}
\affiliation{%
  \institution{Anhui University}
  \department{Key Laboratory of Intelligent Computing and Signal Processing of Ministry of Education}
  \department{School of Computer Science and Technology}
  \city{Hefei}
  \postcode{230601}
  \country{China}
}
\email{zhaoshuzs2002@hotmail.com}
\author{Zhen Duan}
\affiliation{%
  \institution{Anhui University}
  \department{Key Laboratory of Intelligent Computing and Signal Processing of Ministry of Education}
  \department{School of Computer Science and Technology}
  \city{Hefei}
  \postcode{230601}
  \country{China}
}
\email{ycduan@gmail.com}
\author{Jie Chen}
\affiliation{%
  \institution{Anhui University}
  \department{Key Laboratory of Intelligent Computing and Signal Processing of Ministry of Education}
  \department{School of Computer Science and Technology}
  \city{Hefei}
  \postcode{230601}
  \country{China}
}
\email{chenjie200398@163.com}
\author{Yanping Zhang}
\affiliation{%
  \institution{Anhui University}
  \department{Key Laboratory of Intelligent Computing and Signal Processing of Ministry of Education}
  \department{School of Computer Science and Technology}
  \city{Hefei}
  \postcode{230601}
  \country{China}
}
\email{zhangyp2@gmail.com}
\author{Jie Tang}
\affiliation{%
  \institution{Tsinghua University}
  \department{Department of Computer Science and Technology}
  \city{Beijing}
  \postcode{100084}
  \country{China}
}
\email{jery.tang@gmail.com}

\renewcommand{\shortauthors}{Dong Zhang, et al.}

\begin{abstract}
Paper-reviewer recommendation task is of significant academic importance for conference chairs and journal editors. It aims to recommend appropriate experts in a discipline to comment on the quality of papers of others in that discipline. How to effectively and accurately recommend reviewers for the submitted papers is a meaningful and still tough task. Generally, the relationship between a paper and a reviewer often depends on the semantic expressions of them. Creating a more expressive representation can make the peer-review process more robust and less arbitrary. So, the representations of a paper and a reviewer are very important for the paper-reviewer recommendation. Actually, a reviewer or a paper often belongs to multiple research fields, which increases difficulty in paper-reviewer recommendation. In this paper, we propose a \textbf{M}ulti-\textbf{L}abel \textbf{C}lassification method using a \textbf{HIE}rarchical and trans\textbf{PA}rent \textbf{R}epresentation named Hiepar-MLC. Firstly, we introduce a HIErarchical and transPArent Representation (Hiepar) to express the semantic information of the reviewer and the paper. Hiepar is learned from a two-level bidirectional gated recurrent unit based network applying the attention mechanism. It is capable of capturing the two-level hierarchical information (word-sentence-document) and highlighting the elements in reviewers or papers to support the labels. This word-sentence-document information mirrors the hierarchical structure of a reviewer or a paper and captures the exact semantics of them. Then we transform the paper-reviewer recommendation problem into a Multi-Label Classification (MLC) issue, whose multiple research labels exactly guide the learning process. It's flexible that we can select any multi-label classification method to solve the paper-reviewer recommendation problem. Further, we propose a simple \textbf{m}ulti-\textbf{l}abel \textbf{b}ased \textbf{r}eviewer \textbf{a}ssignment (MLBRA) strategy to select the appropriate reviewers. It's interesting that we also explore the paper-reviewer recommendation in the coarse-grained granularity. Extensive experiments on the real-world dataset consisting of the papers in the ACM Digital Library show that MLC-Hiepar achieves better label prediction performance than the existing representation alternatives. Also, with the MLBRA strategy, we show the effectiveness and the feasibility of our transformation from paper-reviewer recommendation to multi-label classification.
\end{abstract}



\maketitle

\section{Introduction}
Paper-reviewer recommendation refers to the automated process for recommending reviewer candidates to perform the paper review. The reviewer recommendation task is very important and meaningful, which enables journal and conference committees to quickly and accurately match papers to reviewers. This core task is not only important in paper reviewing but also is crucial to other scenarios, such as research project selection, question-answerer recommendation and company interviewer selection. It's absolutely a challenge to recommend the appropriate reviewer in these scenarios. Therefore, how to automatically and accurately recommend reviewers for the papers, proposals and questions is a meaningful but still tough task \cite{1}.

More opportunities for computational support should be taken into account in paper-reviewer recommendation process \cite{43}. The problem of paper-reviewer recommendation has been widely studied \cite{2,3,4,5} to reduce the human effort involved in reviewer assignments. Several approaches have been proposed to improve the quality of recommendation and make it perform more accurately. Most of these approaches mainly fall into two categories: retrieval-based methods and matching-based methods \cite{2,6,7,8}. Firstly, the retrieval-based approaches, such as latent semantic index (LSI) \cite{9}, topic model \cite{10,11,12}, etc., mainly focus on the topic relationship between papers and reviewer candidates. Besides, these methods have considered many types of constraints, such as authority, diversity, expertise, availability \cite{2,13,14,15} and the research interests \cite{16} of reviewers. Secondly, the matching-based approaches often compute a matching based on a bipartite graph between papers and reviewer candidates \cite{17,18}. These methods compute the feature-based relevance degree and construct a bipartite graph, and then the right matching can be obtained according to the maximum weight matching based bipartite graph.  

Not only in retrieval-based methods, but in matching-based methods, the feature representation is very important for the relationship or the relevance degree between the paper and the reviewer. The keyword-based feature matching methods often make straightforward matches between the reviewers and the papers based on the number of common keywords. These techniques can result in different matches due to the different granularities of the concepts of keywords. The statistics information, i.e., term frequency-inverse document frequency (TF-IDF) \cite{19} and the topic information produced by the probabilistic latent semantic analysis (PLSA) \cite{7} or latent Dirichlet allocation (LDA) \cite{12} are often used to construct the features from the profiles of papers and reviewers. A system for recommending panels of reviewers for national science foundation (NSF) grant applications, which used a TF-IDF weighted vector space model (VSM) \cite{10} for measuring the association of reviewers with applications. Statistical based methods and topic models are difficult to capture the discriminative semantic contextual information. With the developments of (deep) neural networks, the word2vec methods (CBOW and Skip-gram \cite{20}) learn the word embedding representations for the words of profiles \cite{1}. Nevertheless, these unsupervised embedding features lack the discriminative semantic information of papers and reviewers. This makes learning a more discriminative semantic expression to be a challenge.

\begin{figure}[!htbp]
\centering
\setlength{\abovecaptionskip}{-1ex}
\setlength{\belowcaptionskip}{-0.5ex}
\includegraphics[width=0.6\textwidth]{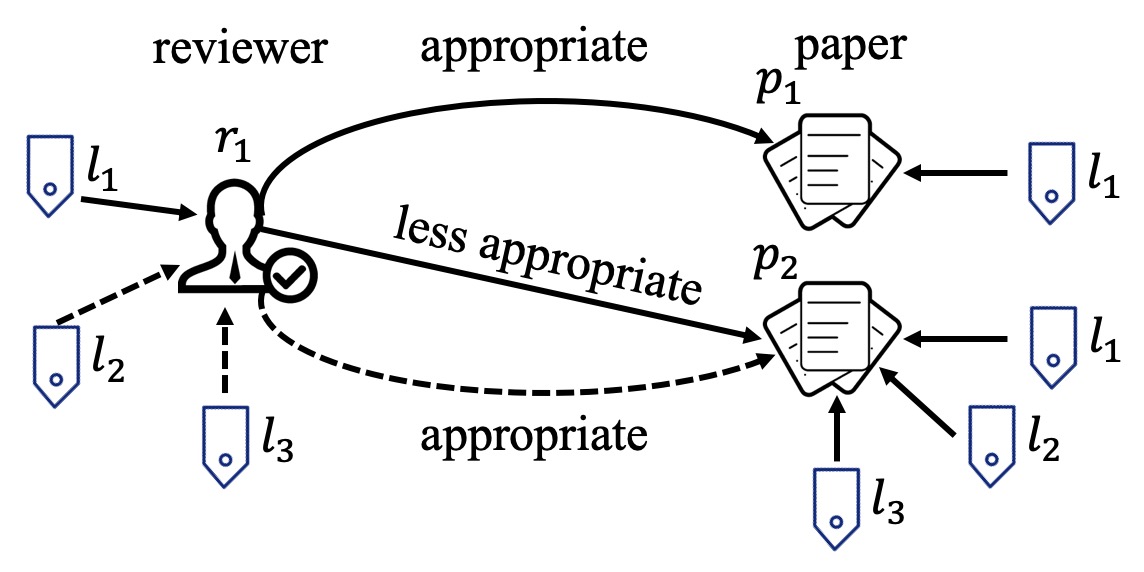}
\caption{Illustration of the scenario that recommending an appropriate reviewer to a paper when they both belong to multiple research fields.}
\label{f1}
\end{figure}

Besides, a paper and a reviewer always belong to multiple research fields, thereby increasing the difficulty in paper-reviewer recommendation. Recommending an appropriate reviewer for a submitted paper is shown in Figure \ref{f1}. $r_1$ is a reviewer who only belongs to the field $l_1$, $p_1$ is a paper that belongs to $l_1$ and $p_2$ belongs to $l_1$, $l_2$, $l_3$. It's appropriate that recommend $r_1$ to the paper $p_1$. However, it's less appropriate to recommend $r_1$ to the paper $p_2$. If the reviewer $r_1$ belongs to $l_1$, $l_2$, $l_3$, it's appropriate to recommend $r_1$ to review the paper $p_2$. Recommending an appropriate reviewer to a paper that belongs to multiple research fields is common but difficult. So it's of great significance to reduce the difficulty in the problem involving multiple research fields in paper-reviewer recommendation.

In this research, we aim to learn more discriminative semantic expressions and reduce the difficulty brought from multiple research fields in paper-reviewer recommendation. To this end, we propose a \textbf{M}ulti-\textbf{L}abel \textbf{C}lassification method using a \textbf{HIE}rarchical and trans\textbf{PA}rent \textbf{R}epresentation named Hiepar-MLC. In fact, there exists a hierarchical word-sentence-document structure in the paper or the reviewer profile, which always consists of the published papers. This hierarchical structure information if beneficial to build semantic expressions for papers and reviewers. So, we introduce a HIErarchical and transPArent Representation (Hiepar), which is learned from a two-level bidirectional gated recurrent unit based neural network applying a two-level attention mechanism. Hiepar attends differentially to more and less important content when constructing the representations for papers and reviewers. Attention mechanism makes the representation transparent and interpretable, highlighting the elements of papers and reviewers, and supporting their labels (research fields). Besides, the task involving multiple research fields is very common but difficult when recommending a more appropriate reviewer to a paper. In particular, the representation of the paper and the reviewer can be more discriminative under the supervision of the research field information. Hiepar builds the more generalized embeddings for the paper and the reviewer, and makes the paper-reviewer recommendation process more robust and less arbitrary. To meet the requirements of recommending the reviewer belonging to multiple research fields, we transform the paper-recommendation problem into a multi-label classification issue. Further, we predict the multiple research labels for a new paper and then recommend a reviewer candidate who has similar labels. It's flexible that we could select any multi-label classification method in multiple labels prediction process. In our method, we select a leading tree-based ensemble algorithm PfastreXML \cite{21}, which is an extreme multi-label (XML) classification method meeting the large label space of any disciplines. Finally, we propose a simple \textbf{m}ulti-\textbf{l}abel \textbf{b}ased \textbf{r}eviewer \textbf{a}ssignment (MLBRA) strategy to select the appropriate reviewers for the paper.

To evaluate our method Hiepar-MLC, we conduct comprehensive experiments on a real-world academic dataset which consists of the papers in the ACM Digital Library. The results and analyses show that our method does build more discriminative semantic expressions for papers and reviewers, and present the transparency from the important elements to the multiple labels. Also, with the MLBRA strategy, we show the effectiveness and the feasibility of our transformation from paper-reviewer recommendation to multi-label classification.

Overall, the main contributions of our paper are as follows:

\begin{itemize}
\item We introduce a hierarchical and transparent representation (Hiepar) for papers and reviewers. It captures the inherent hierarchical structure information to form a semantic expression. Besides, these representations highlight the elements of papers and reviewers, and support their labels (research fields).
\item We transform the paper-reviewer recommendation problem into a multi-label classification issue to meet the requirements of recommending reviewers belonging to multiple research fields. It is flexible that we can select any multi-label classification methods in label prediction process. To the best of our knowledge, our proposed method Hiepar-MLC is the first work that we build the neural network based feature representation in paper-reviewer recommendation, and treat it as a multi-label classification issue.
\item We apply our proposed method on a real-world academic dataset which consists of the papers and the reviewers (authors) in the ACM Digital Library. That presents the potentiality of our method in paper-reviewer recommendation.
\end{itemize}

The remainder of the paper is organized as follows. Section \ref{sec2} presents the related works we investigate in our research. Section \ref{sec3} formally formulates the introduced problem. In Section \ref{sec4}, the details of our proposed method Hiepar-MLC using a hierarchical and transparent representation are presented, and a simple reviewer assignment strategy is introduced. In section \ref{sec5}, extensive experiments and evaluations present the potential of the proposed method in paper-reviewer recommendation. Finally, we summarize our contributions and give the future work in Section \ref{sec6}.

\section{Related works\label{sec2}}
Our work is related to the studies of paper-reviewer recommendation, representation learning and multi-label classification. Therefore, in this section, we briefly review the relevant literatures in these areas.

\subsection{Paper-reviewer recommendation}

Paper-reviewer recommendation plays an important role in academia. It enables journal and conference committees to quickly and accurately match papers to reviewers. Traditional manual reviewer recommendation conducting is fairly involved and labor intensive. It roughly went as follows: the program chair would allocate the papers to the program committee, and send them out for reviews by mails. To make this process effective and labor-saving, many researchers have been studying automated computational support for paper-reviewer recommendation.

In the currently established academic process, paper-reviewer recommendation depends on appropriate assignment to some expert peers for their reviews. With the help of the computational support, paper-reviewer recommendation methods can be mainly divided into two categories: retrieval-based methods and matching-based methods. Retrieval-based methods often transform reviewer recommendation into an information retrieval problem, which treats each submission or manuscript as a query to retrieve the relevant reviewers. Karimzadehgan \cite{7} proposed three general strategies (redundancy removal, reviewer aspect modeling, paper aspect modeling) for reviewer retrieval and studied how to model multiple aspects of reviewer expertise and paper. Karimzadehgan first obtained the text representation for reviewers and papers in topic space, and then retrieved the reviewers with the paper by the similarity. Jaroslaw \cite{22} proposed a recommender system aimed at the selection of reviewers, whereas the recommendations were based on the combination of a cosine similarity between keywords and a full-text index. The keyword features of papers and reviewers exist a lack of semantic information. Also, the manually assigning keywords to define the subject of the paper is subjective. To assign the most appropriate reviewer to the paper, many other constraints are taken into consideration. Liu \cite{13} explored the expert retrieval problem and implemented an automatic paper-reviewer recommendation system that considering the aspects of expertise, authority, and diversity.

The matching-based methods commonly have two steps. First is constructing a bipartite graph and computing the weights of the edges, where the weights are often the similarities between the nodes in the graph. And then an appropriate matching was formulated based on the graph. Shon \cite{8} developed an automatic matching system that matched a research proposal with a reviewer who can evaluate it most effectively, using keywords with fuzzy weights in the corresponding research field. The keywords based matching methods are also influenced by the synonyms and syntactic variants. For example, "SVM" and "support vector machine" are not equivalent in the discrete keyword space, but in fact, they refer to the same underlying concept. To obtain the useful representations of semantically relevant aspects of the paper and the reviewer, the profile-based features are constructed, such as TF-IDF weighted bag-of-words \cite{45}. The matching can be computed with the cosine similarity between the profile-based feature vectors of each paper-reviewer pair. However, the number of the term in profile is often huge, and the low dimensional representation is required. In the work of \cite{10}, the authors employed the statistical topic model to discover expertise and learn the topical components of the papers and the reviewers. Li \cite{23} modeled papers and reviewers with the matching degree by combining collaboration distance and topic similarity. Nevertheless, these profile-based features are obtained in the unsupervised ways, and there still exists a lack of the discriminative semantic information on the papers and the reviewers.

\subsection{Representation learning}

Representation learning of the data makes it easier to extract useful information when building classifiers or other predictors \cite{24}. Creating a more discriminative representation for the paper and the reviewer is crucial to capture the relationship between them. In our study, we mainly focus on the applications of representation learning related to natural language processing. The word space model \cite{25} suggests that a word is characterized by the company it keeps. Gradually, neural network based distributional representation is widely used in text representation learning. CBOW and Skip-gram \cite{20} are two effective models that produce the word embedding which can capture more semantic information. Mikolov \cite{26} represented each document by a dense vector which considered the ordering of the words and the semantics of the words. Successes have been developed in many research domains, which benefit from the embeddings. Kim \cite{27} performed the sentence classification with the convolution neural network (CNN), which made each sentence as an embedding. Palangi \cite{28} modeled the sentence embedding using recurrent neural networks with Long Short-Term Memory (LSTM) cells. Wang \cite{29} predicted the best answerers for new questions by leveraging convolution neural networks in community question answering. Yang \cite{30} proposed a hierarchical representation for the document to perform the text classification task. Tan \cite{31} proposed a simple but effective attention mechanism in question-answer matching task, which constructed better answer representations according to the input question. Gupta \cite{32} studied the problem of scientific paper recommendation through a novel method which aimed to combine multi-modal distributed representations. Recently, a conceptually simple and empirically powerful language representation model, bidirectional encoder representations from transformers (BERT) was proposed. The pre-trained BERT representations can be fine-tuned with just one additional output layer to create state-of-the-art models for a wide range of NLP tasks \cite{46}.

\subsection{Multi-label classification}

Multi-label classification is fundamentally different from the traditional binary or multi-class classification problems which have been intensively studied in the machine learning literature \cite{33}. It is widely used in annotation applications, e.g. image annotation \cite{34}, text categorization \cite{35}. Zheng \cite{36} transformed context prediction to a multi-label classification task, because the contextual conditions can be decomposed into a set of labels. When the label space is an extremely large label set, this multi-label classification problem is an extreme multi-label classification (XML) issue \cite{21}. Tal \cite{37} presented a network consisting of a hierarchical attention bidirectional gated recurrent unit to solve the automatic coding of clinical documentation from large-set diagnosis codes (6500 unique codes). Weston \cite{38} introduced a scalable model for image annotation based upon learning a joint representation of images.

Representations integrating more semantic information for papers and reviewers can improve the quality of the paper-reviewer recommendation. Multi-label classification is appropriate for this task and meets the requirement that recommending a reviewer who belongs to multiple research fields. Hence, in this paper, we aim to transform the paper-reviewer recommendation problem into a multi-label classification issue using a discriminative representation.

\section{ Problem formulation\label{sec3}}

We begin by defining fundamental concepts including reviewers, papers (i.e., submissions or publications), research field labels, and representations of reviewers and papers. Subsequently, we give the formulation of the studied problem. Before continuing, let us briefly introduce the main notation that we will be using in the remainder of the paper; see Table \ref{t1}.

\begin{table}[!hbtp]
\caption{Notations Used in the Article}

    \centering
    \begin{tabular}{p{2cm} p{6cm}}
    \hline
    \textbf{Notation}   &  \textbf{Description} \\\hline
    
    $p, p_i, p_N$   & papers    \\
    $r, r_j, r_M$   & reviewers \\
    $X_{p_i}$   & feature representation of paper $p_i$\\
    $X_{r_j}$   & feature representation of reviewer $r_j$\\
    $y_{p_i}$   & label (research field) vector of paper $p_i$\\
    $y_{r_j}$   & label (research field) vector of reviewer $r_j$\\
    $\textbf{P}$    & paper (submission) set\\
    $\textbf{R}$    & reviewer set\\
    $\textbf{L}$    & label set\\
    $|L|$ & number of labels in $\textbf{L}$\\
    $N$ & number of papers in $\textbf{P}$\\
    $M$ & number of reviewers in $\textbf{R}$\\
    $w_{i_{n_i}}$   & word of paper $p_j$\\
    $w_{j_{n_j}}$   & word of reviewer $r_j$\\
    $n_i$   & number of words of paper $p_i$\\
    $n_j$   & number of words of reviewer $r_j$\\ \hline

    \end{tabular}
    
    \label{t1}

\end{table}

For a reviewer, we always build the profile with a collection of his/her publications. So we assume that the reviewer and the paper are both described as documents consisting of different numbers of words. Given a reviewer who has k publications, we denote it as $r_j= \{w_{j_1},w_{j_2}, ... ,w_{j_{n_j}}\}$. And a paper is denoted as $p_i= \{w_{i_1},w_{i_2}, ... ,w_{i_{n_i}}\}$. Actually, in a scientific domain, there always exists a label set $\textbf{L}$ including a large-number $|L|$ labels. And the paper and the reviewer are often tagged with multiple labels (i.e., belong to multiple research fields). The label vectors of paper $p_i$ and reviewer $r_j$ can be denoted as $y_{p_i}, y_{r_j}\in$ $\left \{ 0,1 \right \}^{|L|}$. We also construct a feature representation for all $p_i$ and $r_j$, respectively denoted as $X_{p_i}$ and $X_{r_j}$. According to the above notations, the paper-reviewer recommendation problem can be formulated as follow.

\textbf{Problem.} Given a reviewer set $\textbf{R}=\{(r_1,y_{r_1}),..., (r_j,y_{r_j}),...,(r_M,y_{r_M})\}$, where $M$ is the number of candidate reviewers and each reviewer  $r_j=\{w_{j_1},w_{j_2},...,w_{j_{n_j}}\}$ is described as $n_j$ words, and $y_{r_j}\in \left \{ 0,1 \right \}^{|L|}$ refers to the label vector of the reviewer $r_j$. In this work, the proposed algorithm is required to transform the paper-reviewer recommendation problem as a multi-label classification issue. So given a paper set $\textbf{P}= \{p_1,p_2,...,p_i,...,p_N \}$ where $p_i = \{w_{i_1},w_{i_2}, ...,w_{i_{n_i}}\}$ is a paper described as $n_i$ words. Based on the reviewer set and the labels, our goal is to predict the paper label $y_{p_i }\in\left \{ 0,1 \right \}^{|L|}$ by the multi-label classifiers. Further, the candidate reviewers can be recommended to a paper if their labels $y_{r_j}$ and $y_{p_i}$ are similar to each other, because of the truth that the reviewer can review the paper when they are in similar research fields.

\section{Proposed method Hiepar-MLC\label{sec4}}
In this section, we present our proposed method to solve the paper-reviewer recommendation problem. More specifically, we first introduce a hierarchical and transparent representation to produce the semantic expressions for papers and reviewers. This representation is learned from a two-level bidirectional gated recurrent unit based network applying attention mechanism. To meet the requirements of recommending reviewers who belong to multiple research fields, we transform the paper-reviewer recommendation into the multi-label classification issue. It's flexible that we can select any multi-label classification methods using the learned representation in paper-reviewer recommendation. Finally, we introduce a simple multi-label based reviewer assignment (MLBRA) strategy to select the appropriate reviewers for the paper.

\subsection{Hiepar: hierarchical and transparent representation}
\begin{figure}[!htbp]
\centering
\includegraphics[width=0.6\textwidth]{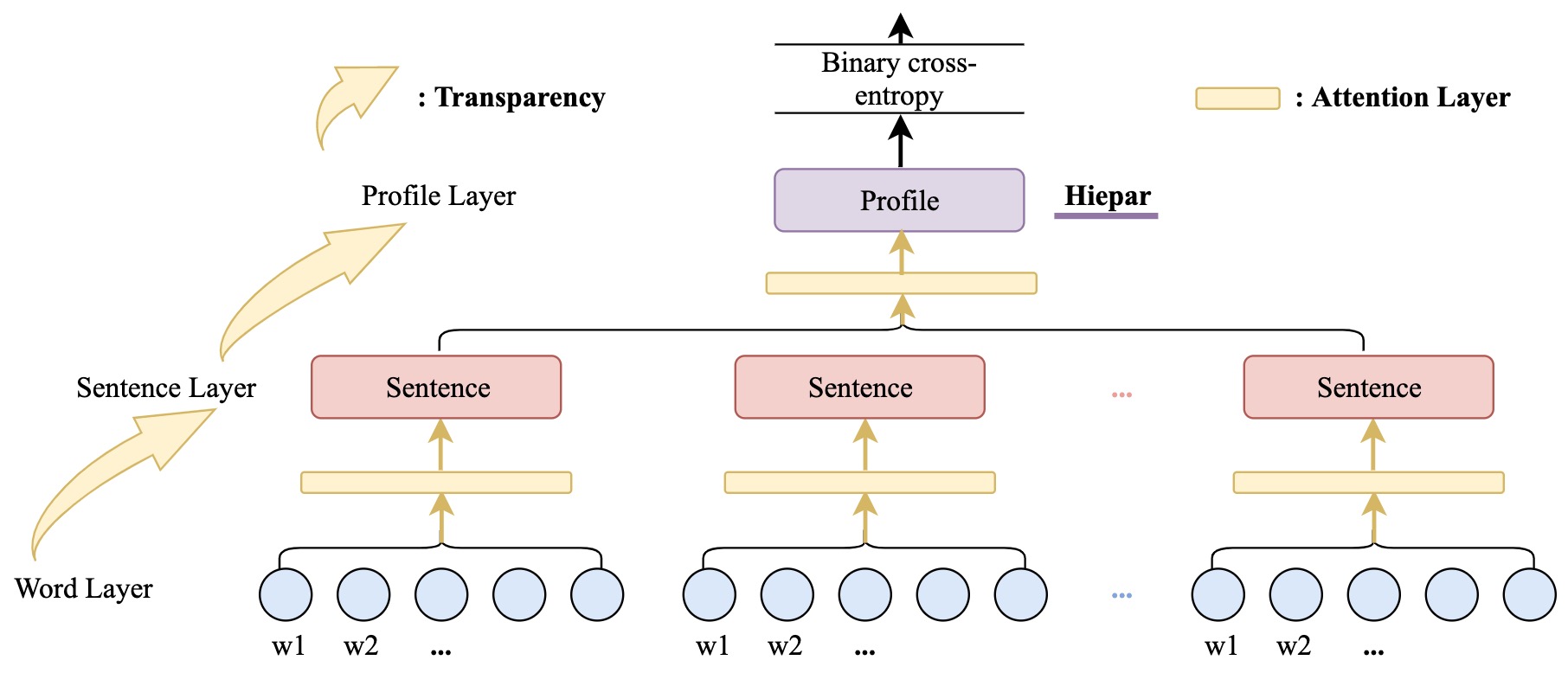}
\caption{The profile representation architecture: the profile layer is our introduced hierarchical and transparent representation.}
\label{f2}
\end{figure}

\textbf{Profile representation.} To make the paper-reviewer recommendation process more robust and less arbitrary, we aim to create the more generalized and inherent embeddings for the paper and the reviewer. So we learn the discriminative expression integrating more semantic information of each reviewer and each paper. In this paper, we introduce a hierarchical neural network with attention mechanism. This hierarchical representation model integrates the word-sentence-document (profile) information into the expressions built from the profiles of the reviewer and the paper. In particular, this inherent hierarchical information makes the representations more generalized. The network consists of two bidirectional GRU \cite{39} layers with the two-level attention mechanism. The first bidirectional GRU operates over tokens and encodes sentences. The second bidirectional GRU encodes the document, applied over all the encoded sentences \cite{37}. From word to sentence and from sentence to document, this model can easily capture the two-level hierarchical word-sentence-document information. The details of Hiepar are shown in Figure \ref{f2}. The representation of the profile layer is the hierarchical and transparent representation (Hiepar).

Assume that the reviewer $r_j$ or the paper $p_i$ is a document that consists of S sentences $s_i$, and every sentence has $T_i$ words. The word $w_{it}$ refers to the $t_{th}$ word in $i_{th}$ sentence. The first bidirectional GRU \cite{39} summarizes contextual information from both directions for words. We can get an annotation for a word $w_{it}$ by concatenating the forward hidden state $h_{forward_{it}}$ and the backward hidden state $h_{backward_{it}}$, i.e., $h_{it}=[h_{forward_{it}},h_{backward_{it}}]$, which captures the information of the whole sentence centered around $w_{it}$.

From the word-sentence view, different words contribute unequally to the representation of the sentence meaning. Therefore, here the attention mechanism helps to extract words that are important to the meaning of the sentence, and then aggregates the representation of these important words to generate a sentence vector.
\begin{equation}
\label{eq1}
u_{it}= \tanh(W_wh_{it}+b_w )
\end{equation}

\begin{equation}
\label{eq2}
weight_{it}=\exp\!(u_{it}^{T}u_w)/\sum_{t}\exp\!(u_{it}^{T}u_w)
\end{equation}

\begin{equation}
\label{eq3}
s_i=\sum_{t}weight_{it}h_{it}
\end{equation}

The word annotation $h_{it}$ is first fed through a perceptron to get $u_{it}$, and then a normalized importance weight $weight_{it}$ can be obtained through a softmax function. Specially, $W_w$, $b_w$ are the parameters of perceptron, $u_w$ is a word context vector and they are all randomly initialized during the training process. After that, the sentence vector $s_i$ is a weighted sum of the word annotations $h_{it}$ based on the weights $weight_{it}$. Finally, the sentence vector $s_i$ integrates the first word-level semantic information, which contributes to the final semantic expressions of reviewers or papers.

From the sentence-document (profile) view, we again feed the given sentence $s_i$ through bidirectional GRU in a similar way to represent the paper or the reviewer profile. We still concatenate the forward hidden state $h_{forward_i}$ and the backward hidden state $h_{backward_i}$, i.e., $h_i=[h_{forward_i},h_{backward_i}]$ which summarizes the information of the all sentences around sentence $s_i$. Different sentences also have different weights in the profile of the reviewer and the paper. We again use the attention mechanism to generate a document vector. Specifically,
\begin{equation}
\label{eq4}
u_i= \tanh(W_sh_i+b_s)
\end{equation}

\begin{equation}
\label{eq5}
weight_{i}=\exp\!(u_iu_s)/\sum_{i}\exp\!(u_{i}^{T}u_s)
\end{equation}

\begin{equation}
\label{eq6}
v=\sum_{i}weight_{i}h_{i}
\end{equation}
where $v$ is the document representation that integrates all the information of all the S sentences. $W_s$, $b_s$ are similarly the parameters of perceptron and $u_s$ is the sentence context vector. They can be randomly initialized and jointly learned during the training process. Eventually, the profile of the reviewer and the paper are regarded as document vectors, which integrate the two-level semantic information.

\textbf{Loss function for paper-reviewer recommendation.} The paper-reviewer recommendation is transformed into a multi-label classification issue. An intuitively reasonable objective for multi-label classification is rank loss \cite{47}, which minimizes the number of mis-ordered pairs of relevant and irrelevant labels. There is a propensity that we aim to tag relevant labels with high scores than irrelevant labels \cite{33}. However, in feedforward neural network architecture, the rank loss has shown to be inferior to binary cross-entropy loss over sigmoid activation when applied to multi-label classification datasets \cite{40}, especially the datasets in the textual domain. Thus, we apply binary cross-entropy loss as the loss function in our neural network architecture to solve the paper-reviewer recommendation. The binary cross-entropy loss function in our profile representation model can be formulated as:
\begin{equation}
\label{eq7}
Loss=-1/M \sum_{i=1}^{M}\sum_{l=1}^{|L|}[y_{r_il} \log \delta({f_{r_il}}) + (1-y_{r_il} )\log {(1-\delta(f_{r_il}))}]
\end{equation}

Where $M$ is the number of reviewers, $f_{r_il}$ is the output of the document vector $v$ through a fully connected layer with $|L|$ units corresponding to the scores assigned to each label, and $\delta(\cdot)$ is the sigmoid function $\sigma(x)$=$\frac{1}{1+e^{-x}}$.

\begin{figure}[!htbp]
\centering
\includegraphics[width=0.6\textwidth]{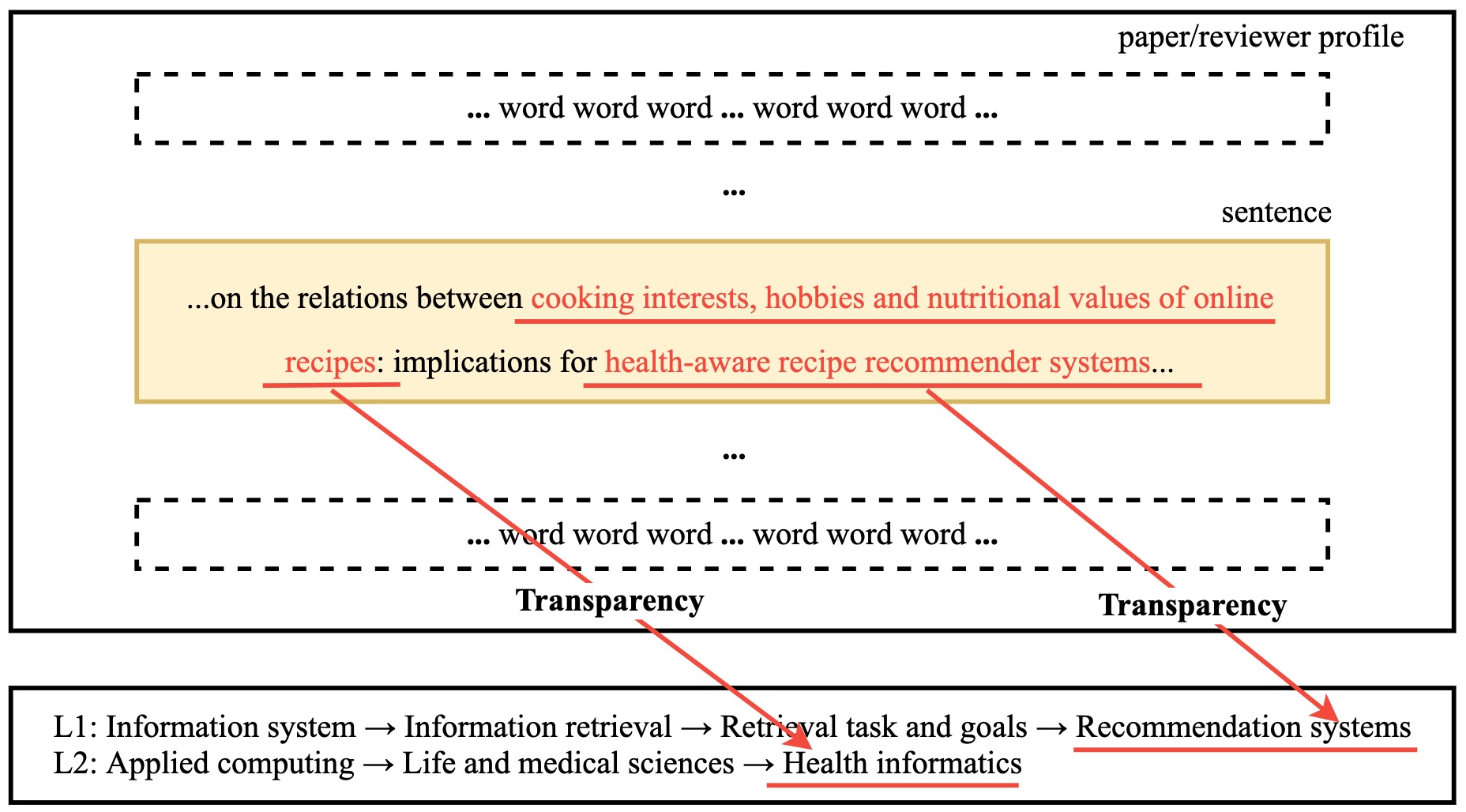}
\caption{The illustration of the transparency. We select a real paper \cite{52} in ACM Digital Library. We focus on some important sentences and the two research fields of this instance paper. It's explicit that the content "cooking interests, hobbies and nutritional values" and "recipe recommender system" support the research field labels "Health informatics" and "Recommender systems" respectively.}
\label{f3}
\end{figure}

\textbf{Transparency property.} As we know, authors always write papers for the specific research fields. To clarify the main idea of the paper, authors often write some sentences that highlight the main topics. In fact, different parts of a document (e.g. paper) possess different proportions of information. Specifically, different sentences in a document are with different weights for each research field label, and different words in a sentence are also with different weights. It's transparent that the important parts with higher weights make the main research topics explicit. This property is called transparency \cite{37}, which can highlight the elements in the documents that explain and support the research labels. Exactly, the two-level attention mechanism can provide full transparency, which is the inherent property of the profiles of the paper and the reviewer, and makes the representation more discriminative.
Here we illustrate the transparency in Figure \ref{f3}.

This hierarchical representation integrating the transparency property here we called hierarchical and transparent representation (Hiepar). We represent papers and reviewers using Hiepar, then we transform the paper-reviewer recommendation problem into a multi-label classification issue in section \ref{4.2}.

\subsection{Multi-label classification for paper-reviewer recommendation\label{4.2}}
\begin{figure}[!htbp]
\centering
\setlength{\belowcaptionskip}{-0.5ex}
\includegraphics[width=0.7\textwidth]{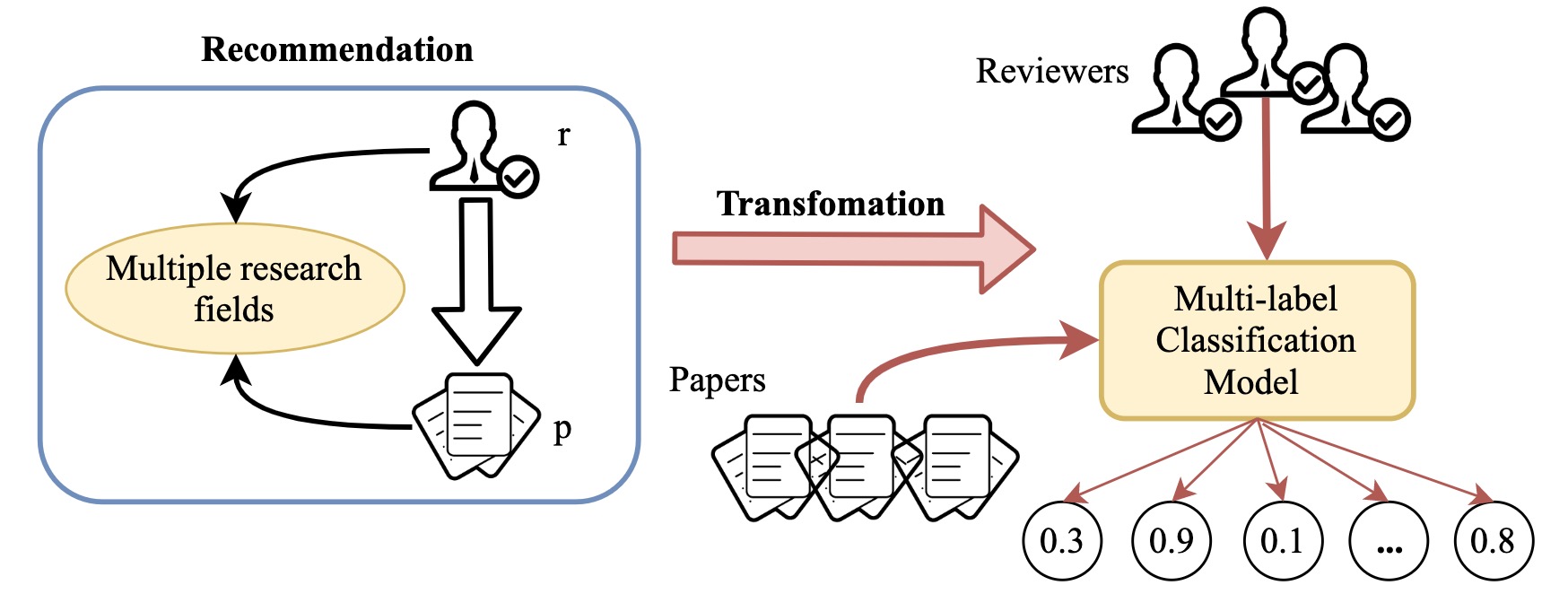}
\caption{ Illustration of transforming the paper-reviewer recommendation problem into the multi-label classification issue.}
\label{f4}
\end{figure}

Considering the difficulty in recommending the reviewers to the paper that belongs to multiple research fields, we transform the paper-reviewer recommendation into the multi-label classification issue. And it's flexible that we can select a multi-label classification method according to the specific scenario. The illustration of the transformation is presented in Figure \ref{f4}. In our multi-label classification process, we try to learn the reviewers of $\textbf{R}$ to train a multi-label classification model. Then we use this model to predict the scores of multiple labels for the papers of $\textbf{P}$. Further, for a submitted paper $p_i$, we can recommend the reviewer $r_j$ who has the same or similar labels.

Actually, in a scientific discipline, such as Computer Science, the paper-reviewer recommendation exists an important characteristic: \textbf{a very large label set} (thousands of unique labels). It's necessary to find the most relevant subset of labels for a paper $p_i$ from an extremely large space of labels. However, the performance of the traditional multi-label classification methods decreases as the number of labels grows and the label frequency distribution becomes skewed \cite{48}. In our paper-reviewer recommendation scenario, we select an extreme multi-label classification method PfastreXML \cite{21}, which is a leading tree-based ensemble algorithm and accurately predicts the scores of relevant labels. 

\subsection{Hiepar-MLC}
To make the representation integrating more semantic information for papers and reviewers, as well as reduce the difficulty in the recommendation in multiple research fields, we propose the method Hiepar-MLC. That introduces a hierarchical and transparent representation and transforms the paper-reviewer recommendation problem into a multi-label classification issue. The details of Hiepar-MLC are presented in Algorithm \ref{a1}.
\begin{equation}
\label{eq8}
    \beta_{ij}=|r_k (\tilde{y}_{p_i})\cap real(y_{r_{j}})|
\end{equation}
\begin{equation}
\label{eq9}
    \beta_{i }^{\ast }=max\left \{\beta_{i1},\beta_{i2},...,\beta_{ij},...,\beta_{iM}\right \}_{j=1}^{M}
\end{equation} 
\begin{equation}
\label{eq10}
    {R_i}'=\left\{...,r_j,...\right\}\quad s.t.\  \beta_{ij}=\beta_{i }^{\ast }\quad j\in\left\{1,2,...,M\right\}
\end{equation}

To recommend the appropriate reviewers to the paper, we introduce a simple multi-label based reviewer assignment strategy (MLBRA). Given a paper $p_i$, $r_k (\tilde{y}_{p_i})$ is the set of the top k relevant labels returned from the multi-label classification model. MLBRA selects the reviewer candidate set ${R_i}'$, where the reviewer has the maximum number $\beta_{i }^{\ast }$ of the similar labels with $r_k (\tilde{y}_{p_i})$ of $p_i$. Finally, we can recommend the reviewer who has more research field labels from ${R_i}'$ to review the paper $p_i$.  Equations \ref{eq8}-\ref{eq10} formulate the MLBRA strategy. For the paper $p_i$, $\beta_{ij}$ is the number of similar labels with the candidate reviewer $r_j$, and $real(y_{r_j})$ is the label set of $r_j$. 
\begin{algorithm}[t]
\KwIn{A reviewer set $\textbf{R}$ = \{($r_1$,$y_{r_1}$), ..., ($r_j$,$y_{r_j}$), ..., ($r_M$,$y_{r_M }$)\}, a paper set $\textbf{P}$ = \{$p_1$, $p_2$, ..., $p_i$, ..., $p_N$\}, and the research field label set $\textbf{L}$.}
\KwOut{The research label $y_{p_i}$ $\in$ $\left \{ 0,1 \right \}^{|L|}$ for each paper $p_i$.}
\textbf{Semantic Representation.} For each $p_i \in P$ and $r_j\in R$, construct the semantic expression respectively denoted as $X_{p_i}$, $X_{r_j}$ $\leftarrow$ $Hiepar$ using Hiepar\;
\textbf{Multi-label Model Training.} On all $X_{r_j}$, train a multi-label classification model $M \leftarrow PfastreXML $  \;
\textbf{Research Label Prediction.} For each $p_i\in P$, predict the $y_{p_i} \leftarrow M$ using the multi-label classification model $M$\;
\textbf{Reviewer Recommendation.} Recommend the reviewer $r_j\in R$ who belongs to similar research label $y_{r_j}$ to the paper label $y_{p_i}$ using MLBRA strategy.
\caption{Hiepar-MLC}
\label{a1}
\end{algorithm}

\section{Experiments and evaluations\label{sec5}}
\subsection{Dataset}
To the best of our knowledge, the real reviewer recommendation process of an academic conference or journal is usually not available. Therefore, we construct a paper-reviewer dataset from the ACM Digital Library. Overall, we obtain 931707 computer science papers including much key information, and we mainly focus on authors (id, name), content (title, abstract), date, and research labels of each paper. Due to the id of the author is unique and we try to build author profile with the content of the published papers. To avoid the impact of the paper quantity of each author, we choose the authors who have 15 papers (publications) at least. We eventually retain 22575 unique authors and we set them as the reviewers in our experiments. Specifically, 13449 papers published in 2017 are regarded as the papers (submissions) which will be reviewed. Notably, we only use the title and abstract information to represent the papers and the profiles of reviewers.

Every paper in ACM Digital Library is tagged at least one label, and these labels are the nodes of CCS taxonomy tree\footnote{https://dl.acm.org/ccs/ccs\_flat.cfm}. In our experiment, we just consider the leaf node of CCS classification tree in the most fine-grained granularity. Therefore, our label space includes 1944 labels totally. The label set of a reviewer is the union of the labels of their own papers.

In our experiments, we set the 22575 reviewers as the training data and set papers in 2017 as the testing data. Firstly, we learn the hiepar for reviewers and papers respectively. Then we transform the paper-reviewer recommendation problem into the multi-label classification issue. For the scenario of 1944-label space, here we select a leading multi-label classification method PfastreXML to train a multi-label model. Finally, we predict the most relevant labels for the papers in the year 2017. Table \ref{t2} reports some descriptive statistics regarding the dataset.

\begin{table}[!hbtp]
\centering
\caption{Statistical description of our dataset}
\label{t2}
\scalebox{1}{
\begin{tabular}{
p{5cm}<{\centering}| p{5cm}<{\centering}}
\hline
Statistics	&Number \\
\hline
labels	&1944\\
\hline
reviewers&	22575\\
\hline
papers	&13449\\
\hline
label per reviewer	&12.88\\
\hline
label per paper	&1.83\\
\hline
\end{tabular}
}
\end{table}

\subsection{Baselines}

We introduce a hierarchical and transparent representation for the paper-reviewer recommendation. For baselines, we select six representation learning models, LDA, CNN, LSTM, Bi-LSTM, Word2vec and BERT \cite{46}. Based on different semantic expressions, we select the same multi-label classification method PfastreXML. For all neural network based baselines, we also select the binary cross-entropy loss function to fit the multi-label issue. Besides, to evaluate the feasibility and the rationality of our transformation that casting the paper-reviewer recommendation problem as the multi-label classification task, we also compare our method with information retrieval methods and topic models, e.g. TFIDF weighted bag-of-words, Word Mover's Distance (WMD) \cite{49}, LSI, LDA and Branch-and Bound Algorithm (BBA) \cite{50}. The details of these baseline methods are followed.

\textbf{CNN.} An adaptation of CNN \cite{27} architecture which leverages three properties that can help improve a machine learning system: sparse interaction, parameter sharing and equivariant representation. This adaptation involves a convolution layer with three kinds of filters (size=3, 4, 5; number=128) and a max-pooling operation. Eventually, it can generate 384-dimensional feature representation.

\textbf{LSTM.} Our LSTM implementation is similar to the one in \cite{41}, which can capture the long-range information and forget unimportant local information. Ours involves only one LSTM layer and the hidden layer size=128. Eventually, it can generate 128-dimensional feature representation.

\textbf{Bi-LSTM.} Bidirectional LSTM \cite{42} uses both the previous and future context by processing the sequence in two directions. Ours involves only one Bi-LSTM layer and the hidden layer size=128. Eventually, it can generate 256-dimensional feature representation.

\textbf{Word2vec.} The average word2vec \cite{20} embedding is used as the feature representation. We use the skip-gram model to generate 100-dimensional features. 

\textbf{BERT} \cite{46}\textbf{.} We use the pre-trained Uncased BERT-Base model \footnote{https://github.com/google-research/bert\#pre-trained-models} (12-layer, 768-hidden, 12-heads, 110M parameters) to obtain the feature representation for the papers and the reviewers.

\textbf{BOW+TFIDF.} TFIDF weighed bag-of-words \cite{45}. The TF-IDF features are computed by multiplying a local component (term frequency) with a global component (inverse document frequency).

\textbf{Latent Semantic Index (LSI)} \cite{9}\textbf{.} A traditional method for automatic indexing and retrieval. LSI is to take advantage of the implicit higher-order structure in the association of terms with documents to improve the detection of relevant documents. We compute the relevance between the papers and the reviewers by the cosine similarity.

\textbf{Latent Dirichlet Allocation (LDA)} \cite{44}\textbf{.} We use LDA to extract the topic vectors of papers and reviewers, and retrieve the appropriate reviewers based on the cosine similarity. We empirically set the number of topics to 100.

\textbf{Word Mover's Distance (WMD)} \cite{49}\textbf{.} WMD is a distance function, which aims to find an optimal matching between words in two documents respectively. We employ it to retrieve the reviewer for the paper. Specially, it uses word2vec to calculate the word embedding of the reviewer and the paper. Then the earth mover's distance is used to calculate the distance between the profiles. We set the dimensional of word embedding to 100.

\textbf{Branch-and-Bound Algorithm (BBA)} \cite{50}\textbf{.} BBA employs LDA to obtain the topic distribution of the reviewer and the paper. Then, BBA performs a group assignment of the reviewers to the papers, considering the relevance of the papers to topics as weights. We set the number of topics to 100.

\subsection{Evaluation metrics}

We transform the paper-reviewer recommendation problem into a multi-label classification issue. We predict the most relevant labels from 1944 labels for the testing papers and we indirectly recommend the candidate reviewers who have a similar label set. Due to the practicality and particularity of our study, we select the recall at top k (Recall@k) and the Normalized Discounted Cumulative Gain (NDCG@k) to evaluate the performance of these methods. The metric of Recall@k is capable of measuring intuitively if the top-k relevant labels are present at the real label set. The metric of NDCG@k illustrates the quality of label ranking. Besides, we adopt the Accuracy to evaluate the performance of the transformation from paper-reviewer recommendation to multi-label classification. For these metrics, the higher value is, the better performance is. These metrics are defined as:
\begin{equation}
\label{eq11}
Recall@k= \frac{1}{\left \| {y_{p_i}} \right \|}\sum_{l\in r_k (\tilde{y}_{p_i})}1^{\{l \;in \;real(y_{p_i})\}}
\end{equation}

\begin{equation}
\label{eq12}
DCG@k= \sum_{l\in r_k (\tilde{y}_{p_i}), n=1,\cdots ,k}
\frac{1^{\{l \;in \;real(y_{p_i})\}}}{\log_2 (n+1)}
\end{equation}

\begin{equation}
\label{eq13}
NDCG@k=\sum_{n = 1,\cdots,k}^{\min(k,\left \| {y_{p_i}} \right \|)}\frac{DCG@k}{\log_2 (n+1)}
\end{equation}

\begin{equation}
\label{eq14}
Accuracy=\frac{1}{N}\sum_{i = 1,\cdots,N}^{}1^{\{real(y_{p_i})\;\cap\;  real(y_{r_{imost}})\; \neq\;  \varnothing  \}}
\end{equation}

Where $r_k(\hat{y}_{p_i})$ is the set of the top $k$ relevant labels of the system prediction, $real(y_{p_i})$ is the true label set of the paper $p_i$, $\left \| {y_{p_i}} \right \|$ counts the number of the true relevant labels, $real(y_{r_imost})$ is the true label set of the most appropriate recommended reviewer for the paper $p_i$, the $N$ is the number of the testing papers and $1^{\left \{ \cdot  \right \}}$ is the indicator function.

\subsection{Experimental Results}
We first make comparisons between our proposed representation method Hiepar and other representation learning approaches (LDA, CNN, LSTM, Bi-LSTM, Word2vec, BERT). Tables \ref{t3} and \ref{t4} demonstrate the performance of Recall@k and NDCG@k of all comparison methods, and the best value is in boldface.

From Table \ref{t3} and Table \ref{t4}, we can see that our Hiepar provides the best performance (the highest Recall@k and NDCG@k scores) on the ACM Digital Library dataset, significantly outperforming the state-of-the-art feature representation method BERT. We attribute such improvement to the effective learning of the multiple research label supervision based on deep neural networks, as well as the attention mechanism to differentiate the importance of words and sentences in reviewers and papers. Although the feature representation generated from the learning processes of Bi-LSTM, LSTM and CNN under the label supervision. From Table \ref{t3} and Table \ref{t4}, we can see that our Hiepar significantly outperforms these three supervised methods. It's obvious that Hiepar can generate a more discriminative feature representation, which integrates the inherent hierarchical structure information.

\begin{table}[!hbtp]
\centering
\caption{Results in Recall@k (R@k, for short) on ACM Digital Library dataset.}
\label{t3}
\scalebox{0.9}{
\begin{tabular}{p{2.5cm}<{\centering}| p{1.2cm}<{\centering}| p{1.2cm}<{\centering}|p{1.2cm}<{\centering} |p{1.2cm}<{\centering}| p{1.2cm}<{\centering}| p{1.2cm}<{\centering}| p{1.2cm}<{\centering}| p{1.2cm}<{\centering}}
\hline
\diagbox{Method}{R@k} & R@1(\%)	&  R@3(\%)&	  R@5(\%)	&  R@7(\%) &R@10(\%)  &R@13(\%) &R@25(\%)   &R@50(\%) \\
\hline 
LDA-MLC &0.11  &0.35  &0.69  &0.93   & 1.41   &3.01    &7.00   &13.41\\
\hline
Bi-LSTM-MLC &0.31  &1.00  &1.90  &2.52   & 3.59   &4.50    &7.53   &12.37\\
\hline
LSTM-MLC  &0.29  &0.94   &1.50  &2.00	&2.90   &3.51    &6.28   &11.22\\
\hline
CNN-MLC  &0.19  &0.71  &1.21  &1.68	&2.30	&3.23    &6.16   &11.15\\
\hline
Word2vec-MLC  &0.14  &0.53  &0.90  &1.32	&2.01	&2.74    &5.97   &11.56\\
\hline
BERT-MLC  &0.55  &1.47  &2.24  &3.22	&4.48	&5.76    &10.04   &16.75\\
\hline
$\bm{Hiepar-MLC}$ & $\bm{0.83}$  &$\bm{2.35}$  &$\bm{3.64}$  &$\bm{5.07}$	&$\bm{6.75}$	&$\bm{8.41}$    &$\bm{13.32}$   &$\bm{20.87}$\\
\hline
\end{tabular}}
\end{table}

\begin{table}[!hbtp]
\centering
\caption{Results in NDCG@k (G@k, for short) on ACM Digital Library dataset.}
\label{t4}
\scalebox{0.9}{
\begin{tabular}{p{2.5cm}<{\centering}| p{1.2cm}<{\centering}| p{1.2cm}<{\centering}|p{1.2cm}<{\centering} |p{1.2cm}<{\centering}| p{1.2cm}<{\centering}| p{1.2cm}<{\centering}| p{1.2cm}<{\centering}| p{1.2cm}<{\centering}}
\hline
\diagbox{Method}{G@k} & G@1(\%)	&  G@3(\%)&	  G@5(\%)	&  G@7(\%) &G@10(\%)  &G@13(\%)    &G@25(\%) &G@50(\%) \\
\hline
LDA-MLC	&0.27	&0.44	&1.09	&1.64	&2.96	&8.54   &24.57	&54.78\\
\hline
Bi-LSTM-MLC	&0.52	&1.48	&3.50	&5.12	&8.41	&11.53   &23.56	&46.33\\
\hline
LSTM-MLC	&0.50	&1.35	&2.60	&3.88	&6.65	&8.71   &19.54	&42.77\\
\hline
CNN-MLC	    &0.36	&1.05	&2.10	&3.32	&5.80	&8.25    &19.80	&43.36\\
\hline
Word2vec-MLC	    &0.30	&0.70	&1.44	&2.51	&4.53	&6.93    &19.60	&45.98\\
\hline
BERT-MLC	    &0.88	&2.01	&3.63	&6.22	&9.95	&14.19    &30.96	&62.30\\
\hline
$\bm{Hiepar-MLC}$	& $\bm{1.38}$	& $\bm{3.24}$	& $\bm{5.98}$	& $\bm{9.72}$	& $\bm{14.74}$	& $\bm{20.21}$   & $\bm{38.95}$	& $\bm{73.66}$\\
\hline
\end{tabular}
}
\end{table}

In particular, each paper in ACM Digital Library has 1.83 labels on average, so we pay more attention to Recall@k and NDCG@k when k=1, 3, 5, 7. Each reviewer has 12.88 labels on average and we try to focus on Recall@k and NDCG@k when k=10, 13, 25, 50. An interesting observation is that our representation Hiepar performs much well than other comparison methods in Recall@k and NDCG@k metrics. Overall, from Table \ref{t3} and Table \ref{t4}, the specific values of these two metrics are generally not high. That means it's challenging to study paper-reviewer recommendation on the ACM Digital Library dataset.

To demonstrate the feasibility and the rationality of our transformation, we also make a comparison between our method with information retrieval methods and topic models (BOW+TFIDF, WMD, LSI, LDA and BBA). We select the most appropriate reviewer according to our introduced simple multi-label based reviewer assignment strategy (MLBRA). For the purpose of consistency, we retrieve the most relevant reviewer to evaluate the performance of these methods. In particular, the average time complexity of solving the WMD optimization problem scales $O(n^{3}logn)$, where $n$ denotes the number of unique words in the profiles \cite{51}. In our experiments, the method WMD failed to perform on the ACM Digital Library dataset due to the very high time complexity. So we construct a small ACM Digital Library dataset, which includes 1000 papers randomly selected from the paper set $\textbf{P}$. From Table \ref{t5} and Table \ref{t6}, our proposed method significantly outperforms other approaches on the two datasets of different sizes. 

\subsection{Case Study and Analysis}
In this section, we provide a case study analysis to demonstrate the improved effectiveness of Hiepar-MLC concerning the label prediction, and illustrate the feasibility and the rationality of our method.
To illustrate the effectiveness of Hiepar-MLC, we show the $top@k$ ($k=\left \|y_{p_i}  \right \|$) label prediction of the paper $p_{155}$. To show the property transparency, we also highlight the important elements of the paper $p_{155}$ (Table \ref{t7:a}) and the reviewer $r_{162}$ (Table \ref{t7:b}), and present the partial content, the real labels. The important words identified by the attention mechanism (Equation \ref{eq2}, where the weights are higher than $0.1$) are highlighted in gold yellow. Specifically, the highlighted elements in the profiles of the paper and the reviewer support the truth labels, which are highlighted in orange yellow. From Table \ref{t7:a}, it's interesting that the meaning of Top3 label $l_{1446}$ is relevant to the truth labels, such as $l_{1277}$ and $l_{1824}$. In fact, the labels $l_{1446}$, $l_{1277}$ and $l_{1824}$ belong to the same coarse-grained category, i.e., \bm{$Software\ and\ its\ engineering$}. To evaluate the performance in a coarse-grained view, we give a comprehensive analysis in next subsection \ref{5.6}.

\begin{table}[!hbtp]
\centering
\caption{Results in Accuracy on ACM Digital Library dataset.}
\label{t5}
\begin{tabular}{p{1.9cm}<{\centering}|
p{2cm}<{\centering}| p{1.15cm}<{\centering}|p{1.15cm}<{\centering} |p{1.15cm}<{\centering}| p{1.15cm}<{\centering}| p{1.15cm}<{\centering}}

\hline
\multicolumn{7}{c}{ACM Digital Library dataset (13449 papers)} \\ 
\hline
Method & BOW+TFIDF& LSI&    LDA&    WMD&    BBA&   \bm{$Ours$}\\
\hline
Accuracy (\%) &5.07  &4.52  &1.84  & $\textbf{-}$&2.65   &\bm{$11.44$}\\
\hline
\end{tabular}

\begin{tablenotes}
\centering
\small
\item[] $\textbf{-}$ denotes the method which failed on ACM Digital Library dataset due to the very high time complexity.\\\bm{$Ours$} denotes \bm{$Hiepar-MLC+MLBRA$}.  
\end{tablenotes}
\end{table}

\begin{table}[!hbtp]
\centering
\caption{Results in Accuracy on ACM Digital Library dataset (small).}
\label{t6}
\begin{tabular}{p{1.9cm}<{\centering}|
p{2cm}<{\centering}| p{1.15cm}<{\centering}|p{1.15cm}<{\centering} |p{1.15cm}<{\centering}| p{1.15cm}<{\centering}| p{1.15cm}<{\centering}}

\hline
\multicolumn{7}{c}{ACM Digital Library dataset (small, 1000 papers)} \\ 
\hline
Method & BOW+TFIDF& LSI&    LDA&    WMD&    BBA&    \bm{$Ours$}\\
\hline
Accuracy (\%) &4.50  &4.80  &2.50  &1.40   &3.24   &\bm{$14.30$}\\
\hline
\end{tabular}

\begin{tablenotes}
\centering
\small
\item[] $\textbf{-}$ denotes the method which failed on ACM Digital Library dataset due to the very high time complexity.\\\bm{$Ours$} denotes \bm{$Hiepar-MLC+MLBRA$}.  
\end{tablenotes}

\end{table}

\begin{table}%
\caption{Case studies of the papers in 2017. The highlighted parts in content and prediction labels show the transparency that the important parts can support the labels. The labels in the bounding box show that the true labels are returned. }
\setlength{\abovecaptionskip}{1pt}
\setlength\belowdisplayskip{1pt}
\label{t7}
\begin{minipage} [c]{1\textwidth}
\subtable[Presentation of paper $p_{155}$.]{
\begin{tabular}{c| p{12cm}<{\centering}}
\hline
paper	& $p_{155}$\\
\hline
content 	&$\cdots$ Can Security Become a Routine? A Study of Organizational Change in an Agile \sethlcolor{gold_yellow}\hl{Software Development} Group Organizational factors influence the success of \sethlcolor{gold_yellow}\hl{security} initiatives in software development. Security audits and \sethlcolor{gold_yellow}\hl{developer} training can \sethlcolor{gold_yellow}\hl{motivate} development teams to adopt security practices, but their interplay with organizational structures and \sethlcolor{gold_yellow}\hl{routines} remains unclear $\cdots$\\
\hline
real labels	& $l_{605}$: Social and professional topics $\rightarrow$ Professional topics$\rightarrow$ Management of computing and information systems$\rightarrow$ Project and people management  \quad \sethlcolor{orange_yellow}\hl{$l_{1277}$: Software and its engineering$\rightarrow$ Software creation and management$\rightarrow$Software development process management} \quad $l_{1824}$: Software and its engineering$\rightarrow$ Software notations and tools$\rightarrow$ Software configuration management and version control systems \\
\hline
prediction	& \fbox{Top1 $l_{1277}$:} Software and its engineering$\rightarrow$ Software creation and management$\rightarrow$ {Software development process management} \quad \fbox{Top2 $l_{1824}$:} Software and its engineering$\rightarrow$ Software notations and tools$\rightarrow$ Software configuration management and version control systems\quad  Top3 $l_{1446}$: Software and its engineering$\rightarrow$ Software creation and management$\rightarrow$ Designing software$\rightarrow$ Software implementation planning$\rightarrow$ Software design techniques \\
\hline
\end{tabular}
\label{t7:a}
}
\end{minipage}

\begin{minipage}[c]{1\textwidth}
\subtable[Presentation of reviewer $r_{162}$.]{
\begin{tabular}{c| p{12cm}<{\centering}}
\hline
reviewer 	& $r_{162}$\\
\hline
content	&$\cdots$by computing correlation coefficients between pairs of human ratings and between human and automatic \sethlcolor{gold_yellow}\hl{ratings}, we found that the automatic scoring of rules based on our novelty measure \sethlcolor{gold_yellow}\hl{correlates} with human judgments about as well as human judgments correlate with one another. text mining semi-\sethlcolor{gold_yellow}\hl{supervised clustering} by seeding joint minimization of power and area in scan \sethlcolor{gold_yellow}\hl{testing} by scan cell reordering $\cdots$and allows the use of a broad range of \sethlcolor{gold_yellow}\hl{clustering distortion} measures, including \sethlcolor{gold_yellow}\hl{bregman} divergences (e.g. euclidean distance and i-divergence) and directional similarity measures (e.g. cosine similarity). we present an algorithm that performs \sethlcolor{gold_yellow}\hl{partitional} semi-\sethlcolor{gold_yellow}\hl{supervised} clustering of data by minimizing an objective function derived from the posterior energy of the \sethlcolor{gold_yellow}\hl{hmrf} model$\cdots$  \\
\hline
real labels	&$l_{782}$: Information systems$\rightarrow$ Data management systems$\rightarrow$ Database management system engines \quad$l_{1223}$: Hardware$\rightarrow$ Hardware validation$\rightarrow$ Functional verification  \quad$l_{1938}$: Information systems$\rightarrow$ Information retrieval  \quad \sethlcolor{orange_yellow}\hl{$l_{1939}$: Information systems$\rightarrow$ Information systems applications$\rightarrow$Data mining}   \\
\hline
\end{tabular}
\label{t7:b}
}
\end{minipage}
\end{table}

In this subsection, we also compare Hiepar-MLC with a common similarity-based method to evaluate the effectiveness of our transformation, which is transforming the paper-reviewer recommendation into the multi-label classification. We use the representation Hiepar as the feature of a paper or a reviewer. Given a paper, we return the most 5 similar candidate reviewers using cosine similarity function.
 
Table \ref{t8} shows the reviewer recommendation results for the given paper $p_{155}$. The most 5 similar reviewers for $p_{155}$ are $r_{17410}$, $r_{5907}$, $r_{8789}$, $r_{4195}$, $r_{1823}$, but only $r_{1823}$ is appropriate to review paper $p_{155}$ to some extent. While the label prediction result for $p_{155}$ using our method is that $p_{155}$ may belong to $l_{1277}$, $l_{1824}$. So we can easily recommend some candidate reviewers who also belong to the research fields $l_{1277}$, $l_{1824}$, such as the reviewers $r_{1348}$, $r_{4603}$, $r_{15377}$.

\begin{table}[!hbtp]
\centering
\caption{reviewer recommendation for paper $p_{155}$ belonging to $l_{605}$, $l_{1277}$, $l_{1824}$}
\label{t8}
\scalebox{0.9}{
\begin{tabular}{p{3cm}<{\centering}|p{3cm}<{\centering}| p{5cm}<{\centering}|p{2cm}<{\centering}}
\hline
method	&recommended reviewers	&reviewers' labels	&appropriate \\
\hline
\multirow{5}*{Hiepar-similarity} &$r_{17410}$ &$l_{182}$, $l_{360}$, $l_{587}$, $l_{613}$, $l_{1413}$	&$\times$ \\
\cline{2-4}
(top 5 reviewers) &$r_{5907}$	&$l_{81}$, $l_{213}$, $l_{1130}$, $l_{1295}$, $l_{1317}$, $l_{1549}$, $l_{1832}$   &$\times$ \\
\cline{2-4}
	&$r_{8789}$	&$l_{385}$, $l_{613}$, $l_{670}$, $l_{1091}$, $l_{1373}$  &$\times$\\
\cline{2-4}
	&$r_{4195}$	&$l_{48}$, $l_{57}$, $l_{246}$, $l_{447}$, $l_{504}$, $l_{842}$, $l_{1032}$, $l_{1226}$, $l_{1426}$, $l_{1460}$, $l_{1614}$, $l_{1666}$ &$\times$ \\
\cline{2-4}
	&$\bm{r_{1823}}$	&$l_{447}$, $\bm{l_{605}}$, $l_{1460}$, $\bm{l_{1824}}$& $\checkmark$ \\
\hline
\multirow{4}*{Hiepar-MLC} & $r_{1348}$	&$l_{93}$, $l_{605}$, $l_{1217}$, $\bm{l_{1277}}$, $\bm{l_{1824}}$	&$\checkmark$\\
\cline{2-4}
	&$r_{4603}$	&$l_{50}$, $l_{787}$, $\bm{l_{1277}}$, $\bm{l_{1824}}$	&$\checkmark$\\
\cline{2-4}
	&$r_{15377}$	&$l_{50}$, $l_{317}$, $l_{605}$, $l_{694}$, $l_{1094}$, $l_{1126}$, $\bm{l_{1277}}$, $\bm{l_{1824}}$ &$\checkmark$\\
\cline{2-4}
	&$\cdots$&$\cdots$&$\cdots$\\
\hline

\end{tabular}
}
\end{table}

\subsection{Recommendation in Coarse-grained Granularity\label{5.6}}
\begin{figure}[htbp]
\centering
\subfigure[ACM Digital Library taxonomy tree]{
\begin{minipage}[t]{0.5\linewidth}
\label{f5:a}
\centering
\includegraphics[width=2.5in]{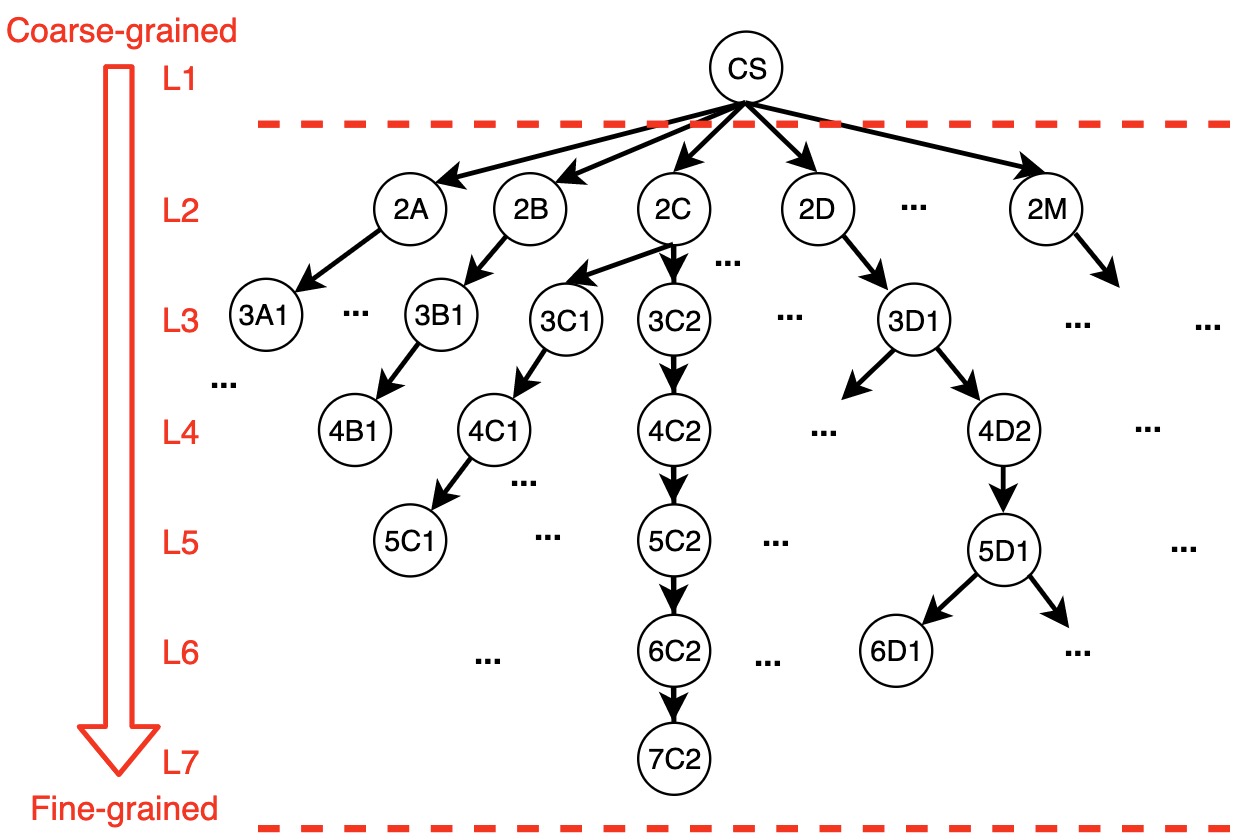}
\end{minipage}%
}%
\subfigure[Label Coarsening]{
\begin{minipage}[t]{0.5\linewidth}
\label{f5:b}
\centering
\includegraphics[width=2.4in]{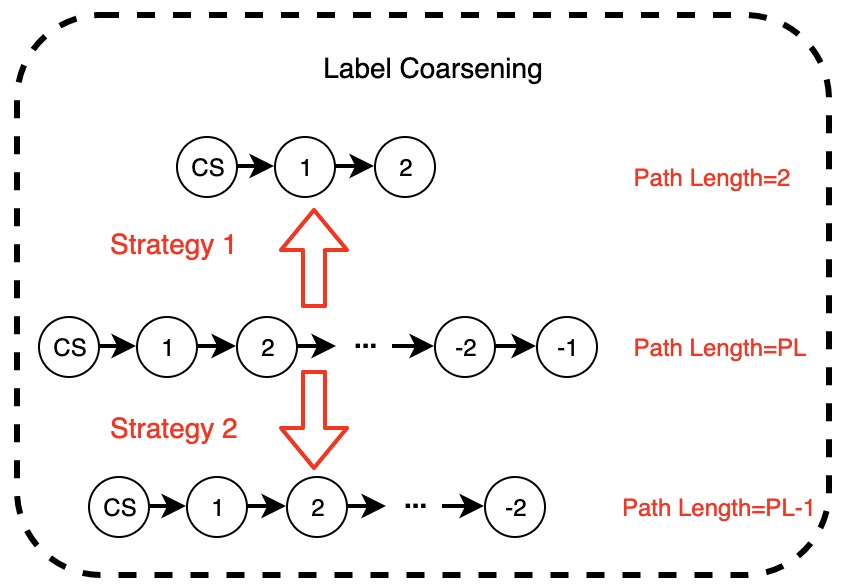}
\end{minipage}%
}%
\centering
\caption{(a) shows the taxonomy tree of the ACM Computing Classification System (CCS). From the first layer (L1) to the last layer (L7), the categories are from the coarse-grained granularity to the fine-grained granularity. The label path $CS\rightarrow2C\rightarrow 3C2\rightarrow 4C2\rightarrow 5C2\rightarrow 6C2\rightarrow 7C2$ indicates a label in the most fine-grained granularity. The nodes indicate the categories of the CCS. For example, the node 5C2 indicates the second kind of category in the $5_{th}$ layer (L5), which is a fine-grained sub-category of the category C (i.e., the node 2C).  (b) shows the label coarsening strategies we use. Strategy 1 aims to preserve the top three successive coarse-grained categories from root to the third node in the current label path. Strategy 2 aims to only remove the most fine-grained category in the current label path. The number indicates the index of the node, where -1 indicates the last node of the label path. }
\label{f5}
\end{figure}

In our ACM Digital Library dataset, the labels are the nodes of the ACM Computing Classification System (CCS). Figure \ref{f5:a} shows the simplified diagram of the CCS taxonomy tree. We use a tree of height $7$ to represent the taxonomy tree of CCS. The root (Computer Science, CS for short) has 13 child nodes, i.e., the 13 coarse-grained categories (the nodes from $2A$ to $2M$ in Figure \ref{f5:a}) in the second layer of the hierarchy. The label of the paper is the path (from the root to a child node or a leaf node) in the tree hierarchy. The paths indicate a series of successive categories from coarse-grained granularity to fine-grained granularity. The path length is in $\{3,4,5,6,7\}$.

\begin{figure}[!htbp]
\centering
\includegraphics[width=0.6\textwidth]{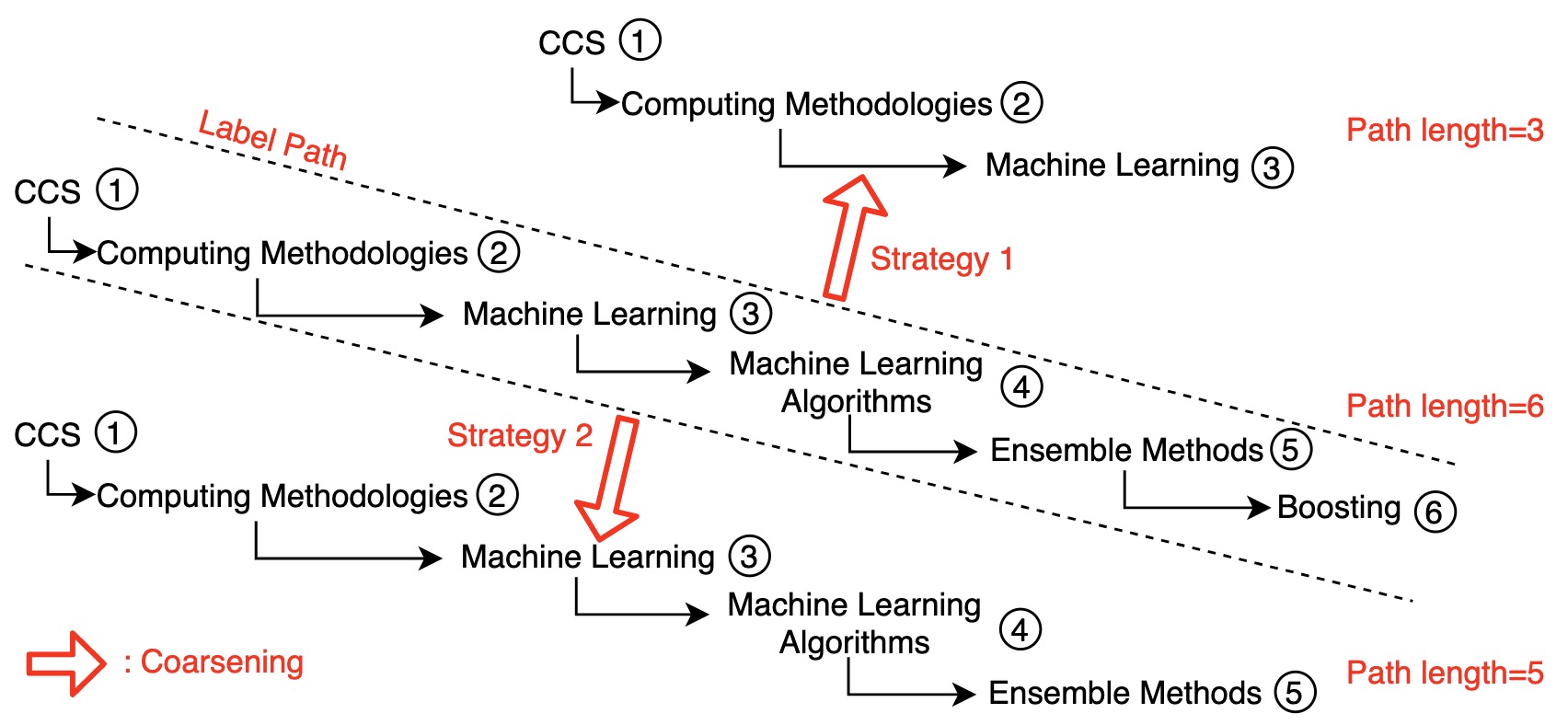}
\caption{The illustration of the label coarsening strategies. The label of length 6 is coarsened as a coarse-grained label of length 3 and 5 under Strategy 1 and Strategy 2, respectively.}
\label{f6}
\end{figure}

To evaluate the performance of our proposed method in the coarse-grained view, we introduce two label coarsening strategies (i.e., Figure \ref{f5:b}). Figure \ref{f6} describes the label coarsening process with a label path instance.

\textbf{Strategy 1:} For a system-predicted label (path), we preserve the top three successive coarse-grained categories from root to the third node in the current path. 

\textbf{Strategy 2:} For a system-predicted label (path), we remove the most fine-grained category (last node) in the current path to form a more coarse-grained label. 

According to these two label coarsening strategies, we again compute the Accuracy to evaluate the performance in the coarse-grained view. The result in Table \ref{t9} shows that our method has the superior Accuracy in paper-reviewer recommendation. Under the \textbf{strategy 1} and the \textbf{strategy 2}, our method obtains the best performance on the ACM Digital Library dataset. That also presents the feasibility and the rationality of our transformation.

\begin{table}[!hbtp]
\centering
\caption{Results in Accuracy on ACM Digital Library dataset using label coarsening}
\label{t9}
\scalebox{0.9}{
\begin{tabular}{p{2cm}<{\centering}|p{2cm}<{\centering}|p{2cm}<{\centering}| p{1.2cm}<{\centering}|p{1.2cm}<{\centering}|p{1.2cm}<{\centering}|p{1.2cm}<{\centering}|p{1.2cm}<{\centering}}
\hline
\multicolumn{8}{c}{ACM Digital Library dataset (13449 papers)} \\
\hline
&Method	&BOW+TFIDF	&LSI    &LDA    &WMD    &BBA    & \bm{$Ours$}\\ \hline
\bm{$Strategy 1$}&Accuracy (\%)	&34.55	&31.61    &17.47    &$\textbf{-}$    &28.96    & \bm{$63.15$}\\
\hline
\bm{$Strategy 2$}&Accuracy (\%)	&34.02	&32.53    &22.88    &$\textbf{-}$    &29.69    & \bm{$55.57$}\\
\hline

\multicolumn{8}{c}{ACM Digital Library dataset (small, 1000 papers)} \\
\hline

\bm{$Strategy 1$}&Accuracy (\%)	&35.70	&34.00    &18.40    &22.10    &28.54    & \bm{$63.00$}\\
\hline
\bm{$Strategy 2$}&Accuracy (\%)	&34.50	&35.00    &24.40    &21.50    &31.22    &\bm{$57.00$} \\
\hline

\end{tabular}
}

\begin{tablenotes}
\centering
\small
\item[] $\textbf{-}$ denotes the method which failed on ACM Digital Library dataset due to the very high time complexity.\\\bm{$Ours$} denotes \bm{$Hiepar-MLC+MLBRA$}.  
\end{tablenotes}
\end{table}

\section{Conclusion\label{sec6}}
In this paper, we mainly investigate the paper-reviewer recommendation problem. We propose a multi-label classification method using a hierarchical and transparent representation (Hiepar-MLC). Due to the particularity of this study, we build a paper-reviewer dataset from ACM Digital Library. Firstly, we construct the profiles of reviewers using their ever publications. Secondly, we learn a representation learning model from reviewers which are set as training data. Notably, we set the papers published in 2017 as the testing data. Based on this model, we obtain the hierarchical and transparent representation of reviewers and the papers. Finally, we solve the paper-reviewer recommendation with a leading multi-label classification method. Furthermore, we introduce a simple multi-label based reviewer assignment (MLBRA) strategy to recommend the candidate reviewers for a paper according to the prediction results of multiple labels. We also study the paper-reviewer recommendation in the coarse-grained granularity. Our experiment results on the ACM Digital Library dataset present that our Hiepar-MLC outperforms the existing representation learning methods and other paper-reviewer recommendation approaches. 

Transforming the paper-reviewer recommendation problem into a multi-label classification issue benefits from using the research label information. In addition, there often exists a hierarchical structure among the multiple research labels of a reviewer or a paper. It's worth extracting the hierarchical information among the labels, which would improve the performance of the label prediction. Also, we would study the paper-reviewer recommendation problem in other scientific disciplines instead of computer science. Actually, the corpora of scientific and technology projects, patents are also valuable to the research on the paper-reviewer recommendation, as well as similar scenarios.

\begin{acks}
This work was partially supported by the National Key Research and Development Program of China (2017YFB1401903), National Natural Science Foundation of China (Grants \#61876001, \#61602003 and \#61673020), the Provincial Natural Science Foundation of Anhui Province (\#1708085QF156), and the Recruitment Project of Anhui University for Academic and Technology Leader.
\end{acks}

\bibliographystyle{ACM-Reference-Format}
\bibliography{Hiepar-MLC.bib}

\end{document}